\documentclass{article}
\usepackage{arxiv}

\usepackage[utf8]{inputenc} % allow utf-8 input
\usepackage[T1]{fontenc}    % use 8-bit T1 fonts
\usepackage{hyperref}       % hyperlinks
\usepackage{url}            % simple URL typesetting
\usepackage{booktabs}       % professional-quality tables
\usepackage{amsfonts}       % blackboard math symbols
\usepackage{nicefrac}       % compact symbols for 1/2, etc.
\usepackage{microtype}      % microtypography
\usepackage{graphicx}
\usepackage[numbers,sort&compress]{natbib}

\title{Recommending research articles to consumers of online vaccination information}

\author{
  Eliza Harrison\\
  Centre for Health Infomatics\\
  Australian Institute of Health Innovation\\
  Macquarie University\\
   \And
  Paige Martin\\
  Centre for Health Infomatics\\
  Australian Institute of Health Innovation\\
  Macquarie University\\
     \And
  Didi Surian\\
  Centre for Health Infomatics\\
  Australian Institute of Health Innovation\\
  Macquarie University\\
     \And
  Adam G. Dunn\\
  Discipline of Biomedical Informatics and Digital Health\\
  Faculty of Medicine and Health\\
  The University of Sydney\\
  \texttt{adam.dunn@sydney.edu.au} \\
}

\begin{document}
\maketitle

\begin{abstract}
Online health communications often provide biased interpretations of evidence and have unreliable links to the source research. We tested the feasibility of a tool for matching webpages to their source evidence. From 207,538 eligible vaccination-related PubMed articles, we evaluated several approaches using 3,573 unique links to webpages from Altmetric. We evaluated methods for ranking the source articles for vaccine-related research described on webpages, comparing simple baseline feature representation and dimensionality reduction approaches to those augmented with canonical correlation analysis (CCA). Performance measures included the median rank of the correct source article; the percentage of webpages for which the source article was correctly ranked first (recall@1); and the percentage ranked within the top 50 candidate articles (recall@50). While augmenting baseline methods using CCA generally improved results, no CCA-based approach outperformed a baseline method, which ranked the correct source article first for over one quarter of webpages and in the top 50 for more than half. Tools to help people identify evidence-based sources for the content they access on vaccination-related webpages are potentially feasible and may support the prevention of bias and misrepresentation of research in news and social media.
\end{abstract}

% keywords can be removed

\section{Background}
The communication of health and medical research online provides a critical resource for the public. More than three-quarters of the UK public report an interest in biomedical research, with 42\% having actively sought out content relating to medical or health research in 2015~\cite{Huskinson2016}. Nearly all searches for health information take place online via search engines~\cite{Castell2014, Huskinson2016,Fox2013,Fox2002}.Internet searches are a common way for people to engage with health research and the communication of health research on news websites and other forums and have the potential to alter health beliefs and decisions~\cite{Weaver2009}.

The communication of health research in news and social media is associated with several challenges. Studies with fewer participants and of lower methodological rigour are more common in news media~\cite{Haneef2017, Selvaraj2014}, and research from authors with conflicts of interest tend to receive more attention in news and social media~\cite{Grundy2018}. As many as half of all news reports manipulate or sensationalise study results to emphasise the benefits of experimental treatments~\cite{Yavchitz2012}.

Despite issues with the reliability of health information online, most people trust what they encounter \cite{Fox2000, Fox2002}, and are inconsistent in their efforts to validate health information using appropriate sources~\cite{Eysenbach2002, Fox2002}, likely because they find it difficult to do so. Where attempts to assess the credibility of health information are made, the visibility and accessibility of sources such as scientific research articles are an important criterion by which users assess the quality of online health communications~\cite{Eysenbach2002, Fox2002}. Individuals are also subject to order-effect biases that impact their perception of the evidence presented by online communications of health research~\cite{Lau2007}, and tend to believe information that aligns with their current knowledge of a health topic~\cite{Fox2002}.

The representation of medical research in the public domain is particularly important in relation to vaccination, where vocal critics actively seek to erode trust in the safety and effectiveness of vaccines and immunisation programs. In 2019, the World Health Organisation listed vaccine hesitancy—the reluctance or refusal to vaccinate—as one of the ten most significant threats to global health \cite{WorldHealthOrganizationWHO2019}. There is a clear risk that the misrepresentation of scientific evidence and amplification of misinformation by social media may be major contributing factors to further outbreaks of these diseases in future \cite{Larson2018}.

The rise of vaccine hesitancy as a global public health issue is in part driven by the increased pervasiveness of anti-vaccination sentiment in search engine results \cite{Kata2012} and the mainstream news media \cite{Larson2011}, as well as the growth of social media as a platform for the provision of a diverse range of information sources to the public \cite{SteffensMarykeS;DunnAdamGandLeask2017}. Discussion of the safety and efficacy of vaccines is a common theme in news reports and low-quality information is common \cite{CooperRobbins2012}. On webpages specifically advocating against vaccination, the majority cite safety risks including illness, damage, or death \cite{Bean2011, Kata2010}.

To be able to identify biases and misrepresentation in the communication of health research online, we need to be able to quickly identify the original source literature for that research. While existing services such as Altmetric (https://www.altmetric.com/) can be used to identify links to scientific source material using Digital Object Identifiers (DOIs), Uniform Resource Locators (URLs), or other identifiers such as PubMed IDs (PMIDs), in most cases these identifiers must be embedded in hyperlinks to enable their tracking. Other media services that offer more complete tracking of media mentions of research tend to be for-profit subscription services that support organisations wanting to keep track of their research outputs. These services are source-centric—they start with a research article and track the media that references it—and may not easily support use cases where a member of the public is interested in accessing the source research that underpins the information on webpages communicating health-related research to the public.

Our aim was to evaluate methods for automatically identifying source literature by recommending articles for webpages communicating vaccination research to the public. To do this, we made use of a large set of reported links between vaccination-related webpages and the scientific literature they reference tracked by Altmetric.

\section{Methods}
\subsection{Study data}
The study data comprised a set of research articles from PubMed linked to a set of webpages via Altmetric. To construct the corpus of research articles from PubMed, we retrieved all articles from PubMed by searching for \textit{“vaccine”}, automatically expanded to include searches for the plural form and \textit{“vaccine”} as a Medical Subject Heading (MeSH) term. Title and abstract text for each article were extracted using the National Center for Biotechnology Information (NCBI) E-Utilities Application Programming Interface (API) (https://www.ncbi.nlm.nih.gov/books/NBK25501/). Any PubMed articles that did not include at least 100 words after concatenating title and abstract were excluded from the analysis, and the remaining 207,538 articles formed the PubMed corpus (Figure 1). The search was conducted in July 2018.

\begin{figure}
\centering
\includegraphics[width=0.9\hsize]{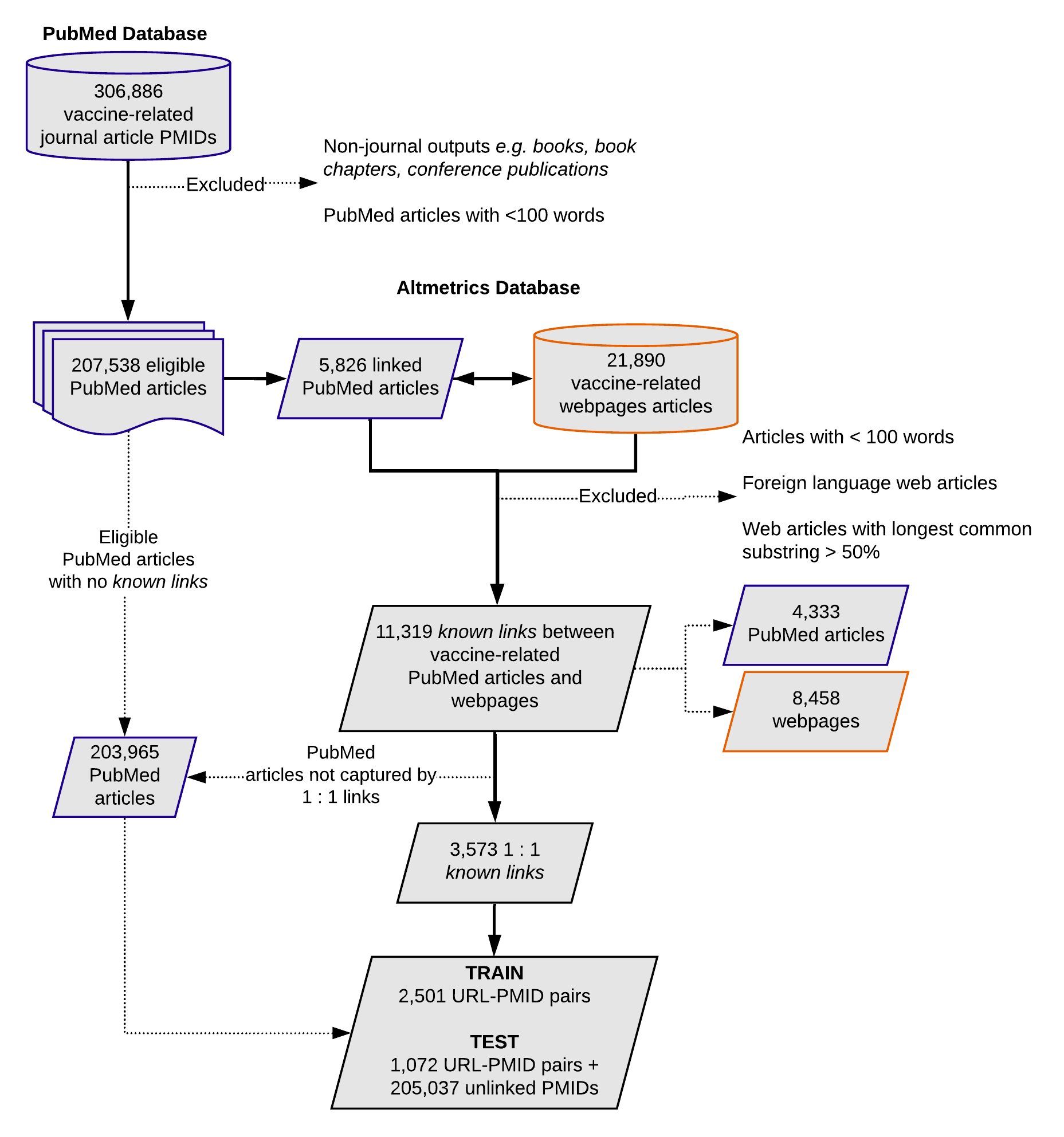}
\caption{The process for the collection and processing of the study datasets. Of the eligible journals articles retrieved from the PubMed database we identified 11,319 distinct known links (URL-PMID pairs) corresponding to research outputs and vaccine-related webpages with links tracked by Altmetric, 3,573 of which were used to train and test the proposed approaches.}
\end{figure}

We then used the Altmetric API to identify the set of research communications that linked to one or more of the articles in the PubMed corpus. We defined research communications to include news articles, blogs and non-social media posts that discuss the outcomes of vaccine-related research. Crawling each URL to access the web articles, contiguous blocks of text from the webpages were concatenated to form the basis of the data used in the following analyses.Text from the set of webpages was accessed in July 2018. Webpages were excluded if they did not include at least 100 words of text, as were any identified as non-English using the Google Code language-detection library (https://code.google.com/p/language-detection/). We also excluded web articles with significant amounts of exact duplicate text. This was common where articles were published on multiple online platforms owned by a single entity, often with only minor changes in title, content, or formatting. To remove these duplicates, we identified webpages for which the longest common substring between any two records linked to a PMID was greater than 50\% of the total length of the longest webpage. We then randomly selected webpages such that no PMID was mapped to any number of similar webpages. Note that after selecting unique examples of linked webpages and research articles, no two webpages had a longest common substring overlap of more than 10\% of the total length.

The resulting dataset included 207,538 research articles, of which 4,333 had known links to one or more of 8,458 distinct webpages (Figure 1). There were 1,934 articles that were referenced on two or more webpages, with one article referenced by 98 distinct webpages. Conversely, there were 1,418 webpages that referenced 2 or more articles, one of which had known links to 68 of the articles in the PubMed corpus. To generate a final set of reported links for which no webpage linked to more than one PubMed article in the final corpus and vice versa, we first selected any article and webpage pairs for which the corresponding PMID and URL were both present only once in the dataset (1:1 links). For each of the remaining articles, we instead selected the linked webpage with the greatest number of words and not yet present in final corpus. This resulted in a final set of 3,573 PMID-URL pairs of individually linked articles and webpages, which we refer to as the \textit{known links} set.

\subsection{Feature extraction and dimensionality reduction}
To generate a term-based vector representation of each of the linked articles and webpages, we pre-processed each document by removing punctuation and words consisting entirely of numeric characters. We then used the remaining words to construct a vocabulary of terms common to both corpora (terms that existed in at least one research article and at least one webpage).

Each article or webpage was then represented as a vector of numeric values based on one of three standard vector representations: \textit{binary}, \textit{term frequency} (TF), and \textit{term frequency-inverse document frequency} (TF-IDF). Binary vectors were generated by recording the presence (value = 1) or absence (value = 0) of vocabulary terms in each document. The TF vector representation was defined as a count of the number of times each word appeared in the document. The TF-IDF score is given by the log-transformed TF value multiplied by the inverse of the log-transformed proportion of documents in which the feature was present. In contrast to term frequency, TF-IDF weights vary depending on how common the term is across the entire corpus, based on the assumption that words appearing more often in fewer documents (like the name of a specific vaccine or the outcomes measured in a research study) are likely to be more informative, while those that appear often across many documents (like “and”, “the”, or “vaccination”) are less informative \cite{SparckJones1972, Ramos2003, Robertson2004}.

In information retrieval methods, sparse representations of documents may be less useful for measuring document similarity or finding documents relevant to a search. This is expected in particular for short documents. To address issues of sparsity, dimensionality reduction methods either remove features that are expected to be less useful or transform the vector space representation into fewer dimensions.

We evaluated the use of two approaches. The first was a simple feature reduction method that uses threshold parameters. Features were removed by applying the maximum document frequency limit of 0.85 to the combined corpora vocabulary. As a result, those terms common to more than 85\% of articles and webpages in the corpus were excluded from the term-based vector representation.

For the second dimensionality reduction approach we used truncated singular value decomposition (T-SVD). T-SVD works in a similar way to singular value decomposition (SVD) by decomposing a matrix into a product of matrices that contain singular vectors and singular values. The singular values can be used to understand the amount of variance in the data captured by the singular vectors. T-SVD allows more efficient computation than SVD since T-SVD approximates the decomposition by only considering a select few components, specified as an argument to the algorithm \cite{Halko2011}.

\subsection{Ranking methods}
We used cosine similarity as a standard measure of similarity between webpages and PubMed articles. For each webpage, we calculated the cosine similarity to all 205,037 articles in the test portion of the final document corpus to produce a ranked list.

We expected that there would be consistent differences between the language style used in article titles and abstracts, compared to that used in online research communications. For example, we expected that communications would replace technical jargon with simpler synonyms. \textit{Canonical correlation analysis} (CCA) \cite{Hotelling1936} is an algorithm designed to identify linear combinations of maximally correlated variables between complex, multivariate datasets. CCA captures and maps the correlations between two sets of variables into a single space, and thus the comparison for ranking can be made using a standard similarity measure. CCA is used to analyse a joint dimensionality reduction across different spaces (e.g., text and images, text and text, etc.) \cite{Menon2015, Rasiwasia2010}. As a result, the CCA approach could be used to learn the alignment between the terms used in the articles and the terms used to describe the same concepts in research communications presented online. To test the CCA approach, we added it as an extra process in the pipeline, using training data to construct a transform (a matrix that may modify the number of features), and then apply that transform to the testing data before calculating the distance (Figure 2).

\begin{figure}[htbp]
\centering
\includegraphics[width=0.9\hsize]{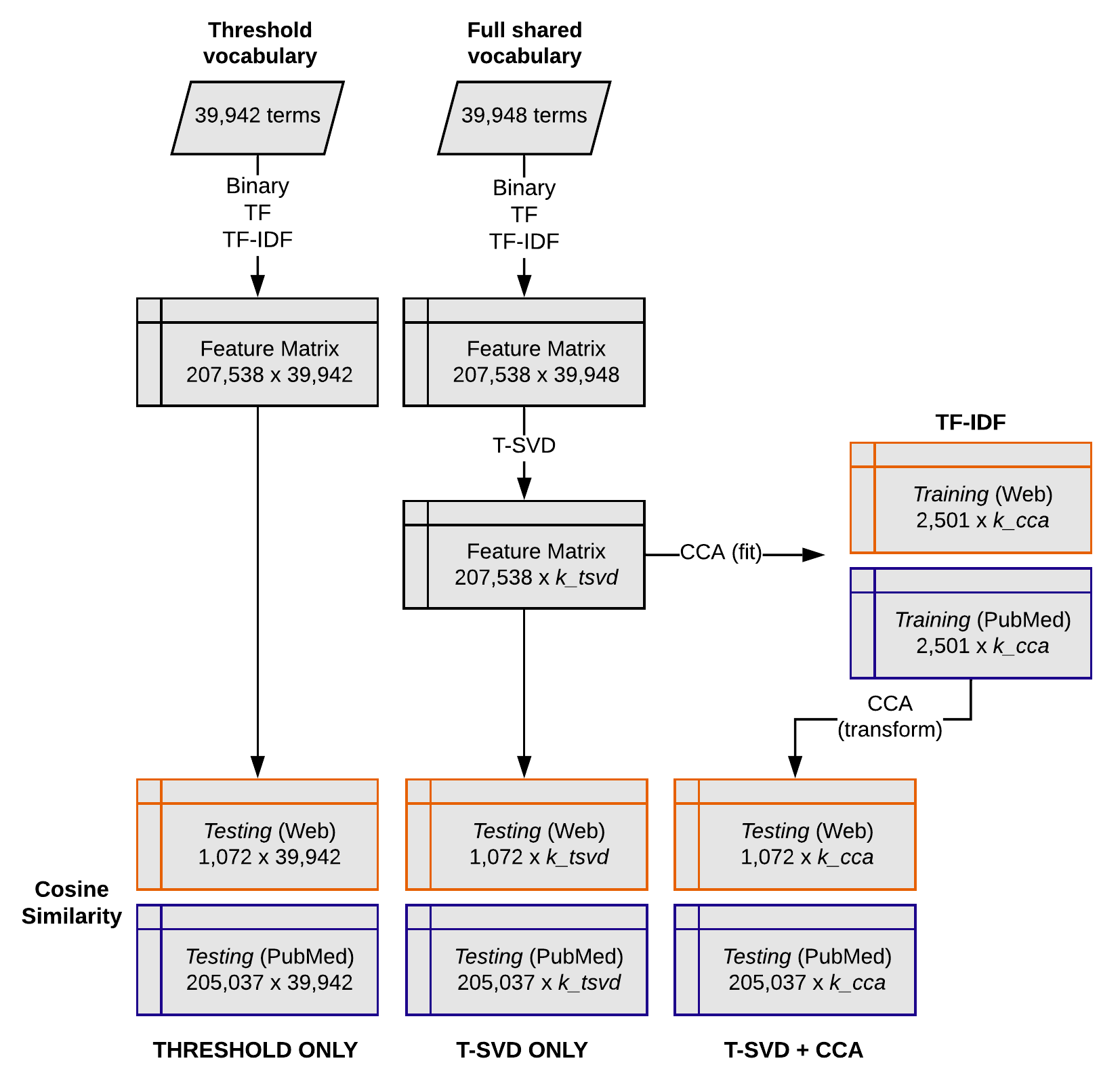}
\caption{We compared three methods of representing terms in using a vocabulary reduced using maximum document frequency parameters (threshold only) or that reduced using T-SVD (T-SVD only), the performance of each was measured by ranking the cosine similarity values between each web and article in the testing set. Also tested was the effect of transforming the best performing feature representation using both T-SVD and CCA on document similarity rankings (T-SVD + CCA).}
\end{figure}

\subsection{Experiments and outcome measures}
While standard document similarity methods typically do not need to be constructed on one set of data and tested on another, the CCA approach learns an alignment between articles and webpages based on a set of training data, and its ability to generalise to unseen data is best tested on a separate dataset. To examine the effect of adding CCA to the pipeline, we constructed training and testing sets by randomly assigning each PMID-URL pair. The resulting training dataset comprised 70\% or 2,501 of the known links, with the remaining 30\% of PMID-URL pairs allocated to the testing set. To replicate the work of searching a large corpus or database for relevant scientific publications, we also added the 203,965 eligible articles not already captured in either the training or testing datasets, resulting in a testing set of 1,072 linked articles and webpages plus the set of 203,965 articles with no linked webpages.

The set of experiments were split into two phases. In the first phase, we examined how differences in the vector space representations might affect the performance of the ranking methods, comparing the binary, TF, and TF-IDF representations in combination with either threshold or T-SVD feature reduction. In the second, we tested the effect of transforming the best performing feature representation using CCA.

The success of each of these systems in correctly linking research articles to the webpages that reference them is indicated by the final rank of the correct PubMed article for each of the 1,072 webpages tested. Based on the similarity between each webpage and source article we calculated the number of PubMed articles a user would be required to read to locate the known links for at least half of all webpages, equivalent to the median rank of the correct source article. As a second metric we determined the number of webpages for which the correct PubMed article was ranked first out of all possible 205,037 articles in the testing set, or the proportion of known links correctly identified by each system (i.e. recall@1). We also calculated the proportion of links ranked within the top 50 PubMed articles in the testing set as an indicator of the capacity of each system to return the correct PubMed article within the first page of query results (i.e. recall@50). Finally, we plotted recall@k for all values between 1 and the total number of PubMed articles to visualise the proportion of known links which can be identified after having read the top k ranked source articles.

All methods and experiments were developed using Python 3.6, the code for which is available on GitHub (https://github.com/evidence-surveillance/web2pubmed).

\section{Results}
Among the 207,538 articles that were returned by the search and met the inclusion criteria for the analysis, 4,333 had one or more links to webpages recorded by Altmetric and were also eligible for inclusion in study analyses. The most popular article was used as source information on 98 webpages, while 22\% (2,535 of 11,319 known links) were used as source information on one webpage (Figure 3). To construct a representative dataset in which no article or webpage was represented more than once, we selected a final set of 3,573 PMID-URL pairs.

\begin{figure}[htbp]
\centering
\includegraphics[width=0.7\hsize]{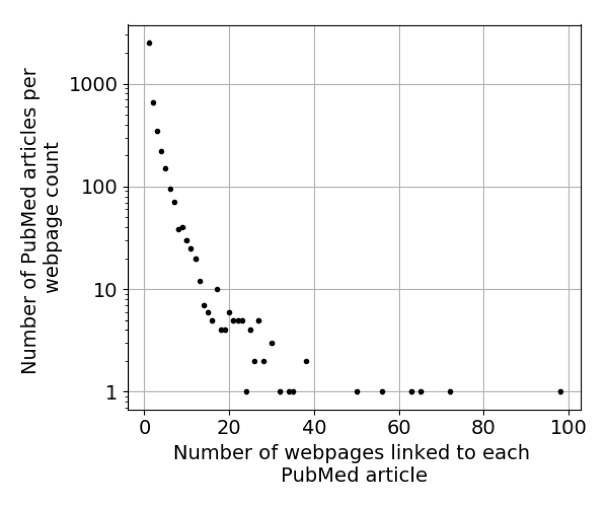}
\caption{The distribution of the number of distinct webpages (URLs) linked to each vaccine-related article retrieved from PubMed (PMIDs).}
\end{figure}

Within this final set of 3,573 articles and webpages with known PMID-URL links and 203,965 additional articles with no known links, we identified 39,948 terms used at least once in both the set of webpages and the set of articles. Where we applied threshold parameters (limiting the vocabulary to exclude terms used in at least 85\% of corpus documents), this vocabulary was reduced to 39,942 terms, representing the greatest number of features used in the following analyses. For experiments instead using the T-SVD method of feature reduction, the number of terms retained in the dataset varied between 100 and 1,600.

Of the methods of representing the text of articles and webpages, we observed that TF-IDF consistently produced the highest performance (Table 1). Regardless of the feature reduction approach used, experiments using the TF-IDF representation of document text outperformed the binary and TF representations.

Of the two feature reduction methods, the threshold approach outperformed the T-SVD approach for all outcome measures (Table 1). However, because the performance improved roughly linearly as the number of T-SVD components was increased, the results suggest that the number of features used may be a more important factor than the choice of feature reduction method. Overall, the highest performance was achieved using TF-IDF to represent the text as term features and the threshold to reduce the number of features. In the testing dataset, the method ranked the correct source article first for more than one in four webpages and placed the correct source article in the top 50 ranked candidate articles for more than half of the webpages.

\begin{table}[htbp]
\begin{center}
\caption{Performance of document similarity methods in a set of 205,037 candidate articles.}
\small
\begin{tabular}{p{4.5cm}ccc}
\hline
\textbf{Feature representation \& reduction methods} & \textbf{Median rank (IQR)} & \textbf{Recall@1} & \textbf{Recall@50}\\
\hline
\textbf{Threshold parameters} \\		
Binary & 238.5 (1-9154)         & 0.251          & 0.417 \\
TF     & 427.5 (5-10075.25)     & 0.188          & 0.368 \\
TF-IDF & \textbf{41 (1-799.25)} & \textbf{0.262} & \textbf{0.515} \\
\hline
\textbf{T-SVD (100 components)} \\
Binary  & 8858 (1198-34252.25)        & 0.049   & 0.097 \\
TF      & 38491.5 (4968.75-104229.25) & 0.046 & 0.077 \\
TF-IDF\textasteriskcentered & 2768 (203.5-24884.5)        & 0.07 & 0.168 \\
\hline
\textbf{T-SVD (200 components)} \\
Binary  & 5522.5 (495-27377.5)    & 0.073 & 0.144 \\
TF      & 36429 (3924.75-99717)   & 0.054 & 0.089  \\
TF-IDF\textasteriskcentered & 1513 (84.75-15572.25)   & 0.097  & 0.225 \\
\hline
\textbf{T-SVD (400 components)} \\
Binary  & 3211.5 (188-21040.25)   & 0.098   & 0.184 \\
TF      & 31220 (2967.25-96203.5) & 0.066  & 0.1  \\
TF-IDF\textasteriskcentered & 720 (36-9674.25)        & 0.126 & 0.276 \\
\hline
\textbf{T-SVD (800 components)} \\
Binary  & 1606 (41.75-15311.75)   & 0.133 & 0.263 \\
TF      & 29421 (2245.25-92871.5) & 0.069  & 0.117 \\
TF-IDF\textasteriskcentered & 385.5 (13-6211.25)      & 0.15 & 0.335 \\
\hline
\textbf{T-SVD (1600 components)} \\
Binary  & 824.5 (9-12704.5)      & 0.173 & 0.331 \\
TF      & 29519.5 (1597.5-93890) & 0.077  & 0.13 \\
TF-IDF\textasteriskcentered & 219 (6-4145.75)        & 0.174 & 0.371 \\
\hline\\
\end{tabular}
\end{center}
\textasteriskcentered  Experiments for which results have also been included in Table 2.\\
IQR: interquartile range; TF: term frequency; TF-IDF: term frequency-inverse document frequency; T-SVD: truncated singular value decomposition.
\end{table}

The addition of CCA was expected to improve the performance of the method by finding an alignment between the terms used in the webpages and articles rather than exact matches between terms. We found that adding CCA to the process improved the performance for experiments where the number of T-SVD components was relatively low (Table 2).
However, as we increased the number of T-SVD components above 400, the improvements gained from adding CCA started to diminish, indicating that the maximum gain in performance from adding CCA was achieved for the experiment that used 400 T-SVD components transformed into 200 feature dimensions by the trained CCA model, where for 38.0\% of the webpages, the correct source article was placed within the top 50 ranked candidates (Figure 4). As the number of feature dimensions used was increased further, the approach then failed because the CCA failed to converge because of the sparsity of the feature space. Overall, the results show that we were able to identify a maximum performance within the parameter space for which the CCA approach could be used, but that none outperformed the simpler approach that used thresholds rather than T-SVD and did not use CCA (Figure 5).

\begin{table}[htbp]
\begin{center}
\caption{Performance of CCA-based alignment methods in a set of 205,037 candidate articles.}
\small
\begin{tabular}{p{4.5cm}ccc}
\hline
\textbf{Method (CCA dimensions)} & \textbf{Median rank (IQR)}      & \textbf{Recall@1}         & \textbf{Recall@50} \\
\hline
\textbf{100 T-SVD components} \\			
No CCA\textasteriskcentered & 2768 (203.5-24884.5)      & 0.07      & 0.168\\
50 & 318 (23-3381)      & 0.099         & 0.319\\
100 & 475 (20-4635.5) & 0.101 & 0.317\\
\hline
\textbf{200 T-SVD components} \\
No CCA\textasteriskcentered & 1513 (84.75-15572.25) & 0.097 & 0.225\\
50 & 322.5 (20-2940) & 0.093 & 0.314\\
100 & 200 (10-1982.75) & 0.133 & 0.358\\
200 & 253.5 (11-4198) & 0.14 & 0.355\\
\hline
\textbf{400 T-SVD components} \\
No CCA\textasteriskcentered & 720 (36-9674.25) & 0.126 & 0.276\\
50 & 575 (60-5055.5) & 0.051 & 0.234\\
100 & 268.5 (15-2696.5) & 0.103 & 0.325\\
200 & \textbf{185.5 (7-2506.75)} & 0.14 & \textbf{0.38}\\
400 & 270 (11-5581) & 0.136 & 0.349\\
\hline
\textbf{800 T-SVD components} \\
No CCA\textasteriskcentered & 385.5 (13-6211.25) & 0.150 & 0.335\\
50 & 3806.5 (279.75-28002.75) & 0.017 & 0.122\\
100 & 1100 (29-15787) & 0.031 & 0.21\\
200 & 409 (27-10816) & 0.084 & 0.296\\
400 & 291.5 (15-9859) & 0.113 & 0.345\\
800 & 1437 (34-34434.75) & 0.075 & 0.272\\
\hline
\textbf{1600 T-SVD components} \\
No CCA\textasteriskcentered & 219 (6-4145.75) & \textbf{0.174} & 0.371\\
50 & 58164.5 (19678-117859.25) & 0.0 & 0.009\\
100 & 47806 (14104.5-110966.5) & 0.001 & 0.023\\
200 & 37414.5 (7236.75-92341.25) & 0.004 & 0.037\\
400 & 30554.5 (3454-91052.25) & 0.005 & 0.06\\
800\textdagger & NA & NA & NA\\
1600\textdagger & NA & NA & NA\\
\hline\\
\end{tabular}
\end{center}
\textasteriskcentered Experiments for which results also appear Table 1.\\
\textdagger Experiments in which the CCA did not converge.\\
IQR: interquartile range; CCA: canonical correlation analysis; t-SVD: truncated singular value decomposition.
\end{table}

\begin{figure}[htbp]
\centering
\includegraphics[width=0.7\hsize]{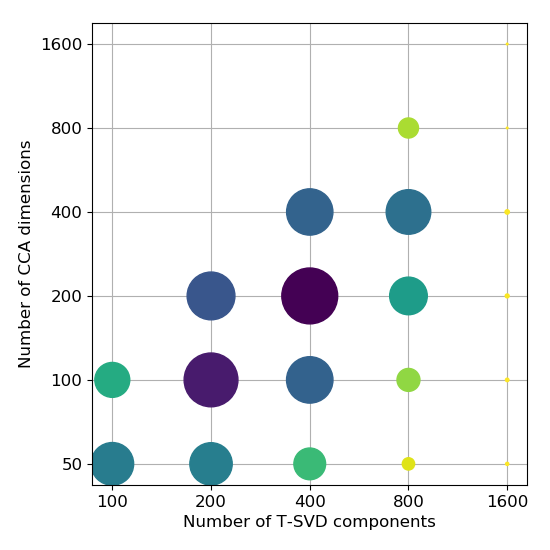}
\caption{A visual comparison of the difference in inverse median rank (circle area) for each of the experiments varying the number of T-SVD components and the number of CCA dimensions. T-SVD experiments where the CCA did not converge are marked with a cross.}
\end{figure}

\begin{figure}[htbp]
\centering
\includegraphics[width=0.9\hsize]{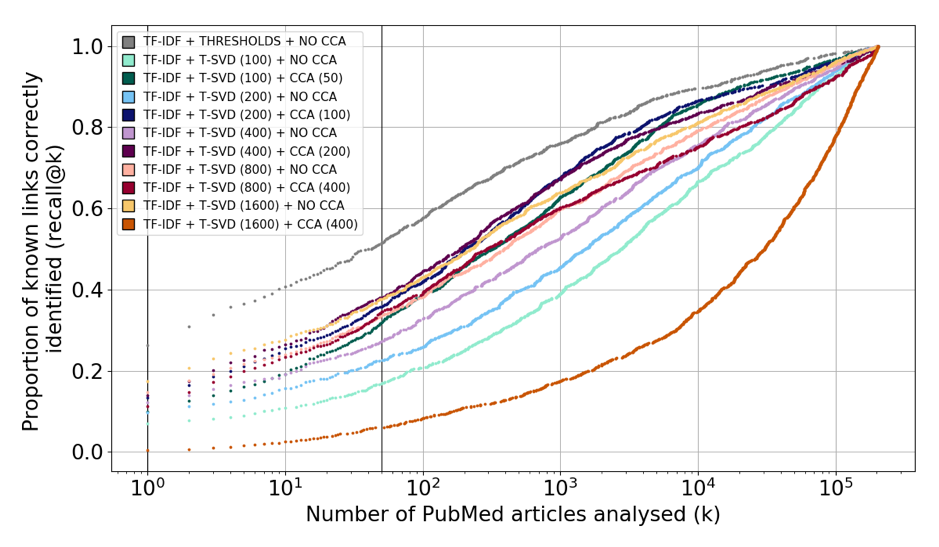}
\caption{The recall@k for experiments comparing the addition of CCA (darker colours) to no CCA (lighter colours), varying the number of T-SVD components and compared to the best performing baseline approach.}
\end{figure}

\section{Discussion}
In this study we evaluated methods that could be used as part of tools to support the identification of missing links between online research communications and the source literature they use. We used vaccination research as an example application domain where there are common problems with bias and misrepresentation in subsequent news and media coverage. We started with the assumptions that many webpages are not reliably connected to the research on which they are based, and that readers may not have the time or expertise to construct a search query to identify relevant articles in bibliographic databases. We tested methods that seek to circumvent the need for expert construction of search queries and instead automatically recommend articles that are likely to be relevant. While the use of a CCA-based approach did not outperform our baseline methods, the results suggest that such tools are likely feasible.

\subsection{Methods for automatic recommendations from text}
We tested two standard information retrieval methods and found that the simpler approach using a TF-IDF representation and a maximum document frequency limit outperformed a more sophisticated approach of transforming the feature space using CCA. While we know of no previous studies that have developed tools for the same purpose, the structure of the problem is common. The combination of TF-IDF and cosine distance has previously been used to identify missing links between trial registrations on ClinicalTrials.gov and articles in PubMed reporting trial results \cite{Dunn2018}. Similarly, the use of TF-IDF has been shown facilitate the detection of similarities between patent documents and scientific publications \cite{Magerman2010}. These results were consistent with ours—increasing the number of SVD components improved the accuracy but the best performance was achieved without the use of SVD.

There are a range of other more complex approaches that could be applied to a problem of this structure: the identification of missing links between two distinct sets of documents that may be matched using similarity of content and a relatively sparse bipartite graph connecting the two sets of documents. These might include alternative feature representations like pre-trained language models, word embedding, or both \cite{Beam2018, Howard2018, Mikolov2013, Peters2018}; as well as other algorithms for recommendation or ranking related to collaborative filtering \cite{Huang2005, Koren2009}, and learning-to-rank methods \cite{Ibrahim2017, Liu2009}.

An expert might take an alternative approach to manually identifying source articles for online research communications, making use of specific information including the names of authors, institutions, or journals. Rule-based approaches that make use of this information may yield improvements. Other similar approaches might make use of the date of publication extracted from webpages and articles in bibliographic databases, under the assumption that online communications of research tend to be reported soon after the research is published.

\subsection{Implications and future applications}
The results indicate that it is likely feasible to build a tool that could be used to help find missing links between health research communications and source literature for the purpose of checking the veracity of the communications and identifying biases. One way to operationalise this type of tool would be to develop browser plugins that automatically augment webpages with a list of recommended relevant peer-reviewed research. Hyperlinks might be added to the terms or phrases that most contribute to the recommendation based on the weights of the terms that contribute to the similarity.

A further application relates to the automatic detection of distortion or bias in research communications. Checklist tools such as QIMR \cite{Zeraatkar2017} or DISCERN \cite{Charnock2004, Charnock1999} are designed to be used to manually evaluate the credibility of health information and health research communications, but little work has been done to use these checklists as the basis for automatically estimating the credibility of webpages \cite{Shah2019}. We know of no studies that have attempted to automatically compare the text of research communications with the abstract or full text of research articles to detect specific differences that might be indicative of misrepresentation of distortion of research conclusions. For example, tools able to identify scenarios where studies of association are written as causation in communications would be of clear benefit, particularly when discussing vaccination \cite{Kata2012, Moran2016}.

Tools extending the work we present here could also be used to help educate non-experts on when it is appropriate to search for source articles when reading research communications online, and to train them on how to construct useful search queries. First, the distances to the top-ranked articles might be suggestive of whether the text on a webpage is based on any form of peer-reviewed research. This could be used to indicate a common practice in anti-vaccine blogs where writers provide circular links within a network of other blogs that are all equally disconnected from clinical evidence. Second, the tool could be used to show users a search query that is automatically generated from the text of research communications for use with bibliographic databases like PubMed, educating users on how to search bibliographic databases for clinical evidence.

\subsection{Limitations}
This study had several limitations. First, while the use of Altmetric helped us to quickly construct a large dataset of reported links, the dataset might be a biased sample of research communications. Communications that include hyperlinks to journal webpages, PubMed, or link to articles using their DOIs may be of higher quality or may be targeted at specialised audiences. Other research communications not using hyperlinks were not included in the dataset, and these may be different to those tracked by Altmetric. Testing the approaches on a more general set of examples before deployment would be necessary. Second, there are a wide range of alternative approaches to feature representation and recommender systems. While we discuss the potential advantages of some of these approaches above, we are at present only able to speculate on which of them are likely to perform best as part of a tool or service aimed at improving the detection of distortion in research communications online. Finally, while vaccination is an important application domain, we did not test what might happen if we had selected a much broader sample of webpages and articles, or if we constructed models specifically designed to find missing links for individual fields or topics of research. It is possible that more general or more specific datasets may influence the performance of the methods we tested.

It is also worth noting that for this dataset, excluding terms not present in both the PubMed and webpage corpora resulted in very few remaining terms were also common to more than 85\% of PubMed articles and webpages, and as such had a minimal impact on the dimensionality of the dataset used for subsequent analyses.

\section{Conclusion}
The results indicate the feasibility of tools designed to support the identification of missing links between health research communications and the scientific literature on which they are based. Such tools have the potential to help people better discern the veracity and quality of what they read online. While standard feature representation and document similarity methods were moderately successful in this task, further investigation is warranted.

%% The bibliography
\bibliographystyle{abbrvnat}
\bibliography{web2pubmedREFS}
\end{document}